\documentclass{article}
\usepackage{meinefortschritte}
\usepackage{amsfonts}
\def\beq{\begin{equation}}                     %
\def\eeq{\end{equation}}                       %
\def\bea{\begin{eqnarray}}                     
\def\eea{\end{eqnarray}}                       
\begin {document}                 
\def\titleline{
%
%
%
D-branes in a marginally deformed WZW model
}
\def\email_speaker{
{\tt 
%
%
forste@th.physik.uni-bonn.de
}}
\def\authors{
%
%
%
%
%
Stefan F\"orste
}
\def\addresses{
%
%
%
%
Physikalisches Institut
der Universit\"at Bonn\\
Nussallee 12,
53115 Bonn,
Germany
}
\def\abstracttext{
%
In this talk we discuss symmetry preserving D-branes on a line of a
marginally deformed $SU(2)$ WZW model. A semiclassical and a quantum
theoretical approach are presented.
       
%
}
\large
\makefront
\noindent This talk is based on the
publications\cite{Forste:2001gn,Hope}. I 
thank Daniel Roggenkamp for a very enjoyable collaboration on\cite{Hope}.

\section{Outline}

After giving a short introduction and motivation I will present the
semiclassical description of D-branes in the marginally deformed
$SU(2)$ model. The second part of the talk addresses the quantum
theoretical description of the D-branes reporting results which will
be derived in more detail in\cite{Hope}. I will finish with some
concluding remarks.

\section{Introduction and Motivation}

Let us first discuss the exact marginal deformations in the case of
closed strings. We know that a (perturbative) string vacuum corresponds
to a two dimensional conformal field theory (CFT). An operator ${\cal
  O}$ is called marginal if it has conformal dimension $(1,1)$. A
typical example of a marginal operator is the product of a chirally
conserved current times an anti-chirally conserved current. An
infinitesimal perturbation of the action by a worldsheet integral over
a marginal operator preserves conformal invariance. Under certain
circumstances\cite{Chaudhuri:1988qb} the infinitesimal perturbation
can be integrated up to a finite exact marginal deformation. This
results in a one parameter family of CFTs, where two adjacent points
on that line are connected by an infinitesimal marginal
perturbation. Hence, exact marginal deformations yield a line in the
space of string vacua and thus help to gain insight into the structure
of the moduli space of string theories. The inclusion of D-branes into
this picture provides information about the moduli space of string
theories in the presence of D-branes. The inclusion of D-branes means
that we investigate CFTs on a worldsheet with boundaries. Such a
theory could also be perturbed by marginal boundary
operators. Deformations resulting from such a perturbation will not be
discussed in the present talk. 

Our plan is to consider as an example the $SU(2)$ WZW model and
deformations thereof. The sigma model description of this class of
CFTs is discussed e.g.\
in\cite{Hassan:1992gi,Giveon:1993ph,Sfetsos:1993ka}.

\section{Semiclassical Description}

Before going to the deformed case let us recall the picture for
symmetry preserving D-branes on $SU(2)$. Two useful parameterizations
of an $SU(2)$ group element are
\begin{eqnarray} g  &=& \cos\! \chi + i \sin \!\chi \cos\!
    \vartheta\; \sigma^1 + i \sin\! \chi 
\sin\!\vartheta \cos\!\varphi \; \sigma^2 + i \sin\! \chi \sin\!
\vartheta \sin\! \varphi \; \sigma^3 \label{para1}\\
 &=& \cos \! x \cos\! \tilde{\theta} -i \sin\!
    x \sin\! \theta\,\sigma^1 +  
i\sin\! x \cos\! \theta\,\sigma^2 + i\cos\! x
\sin\!\tilde{\theta}\, \sigma^3 , \label{para2}\end{eqnarray}
with the parameter ranges $\chi = 0\ldots \pi$, $\vartheta = 0 \ldots
\pi$, $\varphi = 0\ldots \pi$ and $x= 0\ldots \pi/2$, $\theta  =
-\pi\ldots\pi$, $\tilde{\theta} = -\pi\ldots\pi$. The first
parameterization is useful for specifying the location of the D-branes
whereas in the second one the marginal deformations look simple. The
corresponding target space metric is the Cartan-Killing metric
 \begin{eqnarray*}
ds^2 & = &  k\alpha^\prime \left\{ d\chi^2 + \sin^2 \!\chi\, \left(
    d\vartheta^2 
    +\sin^2\! \vartheta\, d\varphi ^2\right)\right\} \nonumber \\
& = & k\alpha^\prime \left\{ dx^2 +\sin^2\! x\, d\theta^2 + \cos^2\! x
    \, d\tilde\theta^2 \right\},
\end{eqnarray*}
leading to an $S^3$ geometry. The radius of this three sphere is
quantized, $k\in {\mathbb N}$ \cite{Witten:ar}.

The symmetry preserving D-branes were described
in\cite{Alekseev:1998mc}\footnote{A semiclassical discussion of
  symmetry breaking branes can be found in\cite{Behrndt:1993fh}.} as
conjugacy classes of a 
  fixed group 
element. These are spherical branes breaking the $SU(2)\times SU(2)$
Kac-Moody to a diagonal $SU(2)$ symmetry. In terms of isometries the
$SO(4)$ of $S^3$ is broken to an $SO(3)$ acting as rotations along the
branes. 
Further, a topological argument let the authors of\cite{Alekseev:1998mc}
to a 
quantization condition on the position of the
D-branes. In the 
parameterization (\ref{para1}) the position of the brane is specified
by fixing the value of $\chi$ and the quantization condition reads
\begin{equation}\label{quantclass}
\chi \in \frac{\pi {\mathbb Z}}{k}.
\end{equation}
Later in\cite{Felder:1999ka} the position of the D-branes was studied
by probing the geometry with gravitons and it was found that 
(\ref{quantclass}) is quantum corrected to
\begin{equation}
\chi \in \frac{\pi(2j +1)}{k+2} \,\,\, ,\,\,\, j =0,\frac{1}{2}, 1,
\ldots ,\frac{k}{2} .
\end{equation}
We will come back to this method of investigating the D-brane geometry
later. In the present section we consider the large $k$ limit in which
the quantization conditions coincide apart from the zero dimensional
branes at the poles. We will not include branes at the poles into the
discussion.   

At first sight this picture seems counterintuitive since the
two-dimensional tension-full branes wrapping contractable cycles are
expected to shrink to zero volume. The mechanism responsible for
stabilizing the two dimensional branes has been given
in\cite{Bachas:2000ik, Pawelczyk:2000ah} (see
also\cite{Bordalo:2001ec} for more general groups) : The quantization
condition 
on the position of the D-brane (\ref{quantclass}) corresponds to a
quantization condition on the F-flux through the D-brane.

The metric for the deformed $SU(2)$ model can be found in
e.g.\cite{Giveon:1993ph}. In the coordinates (\ref{para2}) it reads
\begin{equation}
ds^2/k\alpha^\prime  = dx^2 +\frac{ \sin^2x\, d\theta^2 + R^2\cos^2x\, 
  d\tilde{\theta}^2}{\cos^2 x + R^2 \sin^2 x} ,
\end{equation}
where the deformation parameter $R$ takes values in $(0,\infty
)$. The value $R=1$ corresponds to the undeformed model. In the limit
$R\to\infty$ ($R\to 0$) the $\theta$ ($\tilde{\theta}$) coordinates
decouple and the limiting geometries are $SU(2)/U(1)$ times a free
boson on a circle of vanishing radius (which can be T-dualized to the
real line). The geometry of $SU(2)/U(1)$ can be conformaly
compactified to a disk\cite{Maldacena:2001ky}. Away from $R=1$ the
$SU(2)\times SU(2)$ Kac-Moody symmetry is broken to a $U(1)\times
U(1)$ symmetry. The corresponding chirally and anti-chirally conserved
currents combine to the marginal operator taking the model from a
point $R$ to an infinitesimal close point $R+\delta R$.
The fate of the symmetry preserving D-branes was studied
in\cite{Forste:2001gn} where the following rules were imposed. The
D-branes should break the $U(1)\times U(1)$ to a residual $U(1)$ at
generic $R$ and for $R=1$ they should be identical to the previously
discussed symmetry preserving branes on $SU(2)$. Further an F-flux
quantization condition should be satisfied. This lead to the result
that the position of the D-branes expressed in the coordinates
(\ref{para2}) does not change with $R$. In the remaining part of the
talk we want to confirm this result by a quantum theoretical
discussion.
  
\section{Quantum Theoretical Description}

In order to go beyond the semiclassical treatment of the previous
section we should construct the boundary states for the symmetry
preserving D-branes on the line of deformed models. What we will do is
to give a set of boundary states on the family of deformed models and
claim that a certain subset of those boundary states corresponds to
the symmetry preserving D-branes. Afterwards we will justify our claim
by showing that in the large $k$ limit the geometries are identical. 
In order to construct the boundary states on the deformed models an
alternative description of the deformation turns out to be
useful. This description was proposed in\cite{Yang:1988bi}. The
statement is that for arbitrary $R$ there is an
identification 
\begin{equation} 
\mbox{Deformed Model} = \left( \mbox{pf}_k \times
  u(1)_{\sqrt{k}R}\right)/{\mathbb Z}_k .
\label{yang}
\end{equation}
Let us briefly sketch the ingredients of this
construction. First the parafermions ($\mbox{pf}_k$) posses a
${\mathbb Z}_k\times {\mathbb Z}_k$ symmetry with the currents
$$\psi_l\left( z\right) \,\mbox{and}\,\,\, \psi^\dagger _l\left(
  z\right)$$
whose charges are $\left( l , 0\right)$ and $(-l,0)$, respectively
($l=0,\ldots, k-1$). One example for an OPE in $\mbox{pf}_k$ is
\begin{equation}
\psi_l\left( z\right)\psi_p\left( 0\right) \sim z^{-lp/k} \left(
  \psi_{l+p}\left( 0\right) + {\cal O}\left( z\right)\right) .
\end{equation}
Since moving $\psi_l$ once around $\psi_p$ yields a phase which is not
necessarily $\pm 1$ the model is called parafermion model. 

The  other ingredient is a free compact boson
$$\phi\left( z ,\bar{z}\right) = \phi\left( z\right) +
\bar{\phi}\left(\bar{z}\right) \,\,\, , \,\,\, \left< \phi\left(
    z\right) \phi\left( 0\right)\right> = -2\log z .$$
That the above model is for $R=1$ equivalent to the $SU(2)$ model has
been pointed out in\cite{Gepner:1986hr}. The $SU(2)$ currents can be
represented as follows,
\begin{eqnarray*}
J^+\left( z\right) &=& \sqrt{k}\, \psi_1\left( z\right)
: e^{\frac{i\phi\left(z\right)}{\sqrt{k}}}: \nonumber\\
J^-\left( z\right) &=& \sqrt{k}\, \psi^\dagger _1\left( z\right)
:e^{\frac{-i\phi\left(z\right)}{\sqrt{k}}}: \nonumber\\
J^3\left( z\right) &=& \sqrt{k}\, \partial_z \phi\left( z\right) ,
\end{eqnarray*}
which can be checked by comparing the OPE's. The right hand sides are
invariant under a ${\mathbb Z}_k$ transformation acting on the
parafermions and simultaneously shifting the free boson. Thus the
${\mathbb Z}_k$ orbifold group in (\ref{yang}) is a diagonal subgroup
of the ${\mathbb Z}_k \times {\mathbb Z}_k$ symmetry in $\mbox{pf}_k$
combined with a discrete shift along the $u(1)$. The deformed models
are now reached by changing the size of the circle on which the boson
$\phi$ lives or more generally spoken deforming the Cartan torus of
the group. This can be confirmed by matching the zero mode spectrum of
the orbifold (including twisted sectors) with the eigenvalues of the
Laplacian (with non-trivial dilaton, see e.g.\cite{Maldacena:2001ky})
appearing in the effective closed string action.

The construction of boundary states in the description (\ref{yang})
turns out to be rather simple since all the input one needs can be
found in the literature. The boundary states in $\mbox{pf}_k$ have
been constructed in\cite{Maldacena:2001ky}\footnote{
For discussions of D-branes on cosets see
also\cite{Gawedzki:2001ye,Parvizi:2001xe,Elitzur:2001qd,Fredenhagen:2002qn,Ishikawa:2001zu,Kubota:2001ai,Walton:2002db}. 
}
and the boundary states for
a free boson on a circle are simple. A boundary state on the orbifold
is obtained by tensoring these two and adding all ${\mathbb Z}_k$
images. Our claim is that the symmetry preserving D-branes of the
previous section arise by combining an A-type boundary state of
$\mbox{pf}_k$ with
Dirichlet-type boundary state of the free boson (with a certain
position\footnote{For general positions $x$ the D-brane is localized
  on a twisted conjugacy class\cite{Stanciu:1999id}.}). The A-type
  boundary state of the parafermion theory reads 
\begin{equation}\label{apfk}
\left| B_{(j,n)} ^k\right> =\sum_{\left(j^\prime ,n^\prime\right)}
\frac{S^k_{(j,n), (j^\prime ,n^\prime)}}{\sqrt{S^k_{(0,0)(j^\prime ,
      n^\prime)}}} \left|\left. \left( j^\prime ,
      n^\prime\right)\right>\right>, 
\end{equation}
where the notation follows\cite{Maldacena:2001ky}. In particular the
pair $(j,n)$ labels the highest weight states and the $S$ matrix
arises in modular transformations of the closed string partition
function. The action of an $l$th order element $\ell$ of the orbifold group
${\mathbb Z}_k$ follows from its action on the Ishibashi
state\cite{Ishibashi:1988kg} 
\begin{equation}
\ell\left|\left. \left( j^\prime , n^\prime\right)\right>\right> = 
\exp\left\{ 2\pi i ln^\prime/k\right\}\left|\left. \left( j^\prime ,
      n^\prime\right)\right>\right> .  
\end{equation} 
The Dirichlet-type boundary state on $u(1)_{\sqrt{k}R}$ reads
\begin{equation} \label{dletb}
\left| D^{\sqrt{k}R}\left( x\right)\right> = \sum_{p\in {\mathbb Z}}
\frac{e^{2\pi ipx/k}}{\sqrt{\sqrt{k}R}}
\left.\left| \left( p,0\right)\right>\right>_D ,
\end{equation}
where the momentum number $p$ labels the highest weight states and $x$
specifies the position of the D-brane. We see that $R$ enters only as
a normalization. The ${\mathbb Z}_k$ action on the Ishibashi state in
(\ref{dletb}) is
\begin{equation}
\ell \left.\left| \left( p,0\right)\right>\right>_D = \exp\left\{ -2\pi i
  lp/k\right\} \left.\left| \left( p,0\right)\right>\right>_D .
\end{equation}

The sum over the ${\mathbb Z}_k$ images of the tensor
product is non-vanishing only if the momentum number of the Dirichlet
Ishibashi state coincides with $n^\prime$ in the $\mbox{pf}_k$ Ishibashi
state which is labeled by the pair $(j^\prime ,n^\prime )$ as  
in\cite{Maldacena:2001ky} (see also (\ref{apfk})). For a certain value
for the position of the 
Dirichlet boundary state the boundary state on the orbifold takes the
form of an $SU(2)$ Cardy state\cite{Cardy:ir} belonging to symmetry
preserving branes on $SU(2)$.  

It remains to show that these boundary states are indeed the ones we
found in the semiclassical considerations, before. To this end, we
need to derive the geometry of the D-brane given by the boundary
state. The problem of deriving the geometry from a given boundary
state has been analyzed in\cite{DiVecchia:1997pr} for flat target
spaces and in\cite{Felder:1999ka} for group manifolds. 
In both cases one identifies the D-brane geometry as the set of points
on which the overlap between a closed string graviton state and the
boundary state does not vanish (for $k\to\infty$).
For our purpose the discussion in\cite{Felder:1999ka} can be used in a
straightforward 
way. As we already mentioned the boundary state in the deformed model
takes the form of a symmetry preserving $SU(2)$ Cardy state. The other
information one needs is that the form of the eigenfunctions of
$$ \frac{e^{2\Phi}}{\sqrt{G}}\partial_\mu\left( e^{-2\Phi}\sqrt{G}
  G^{\mu\nu}\partial_\nu\right) $$
does not change with $R$ (here $\Phi$ and $G$ denote the dilaton and
metric in the deformed model). This implies that the highest weight
contribution to the closed string graviton state does not change its
form under the deformation. Using this and going through the 
discussion 
of\cite{Felder:1999ka} one finds that the D-branes are localized on
conjugacy classes of a fixed group element. This means that the
D-brane position expressed in the coordinates of the undeformed model
does not change under the deformation. 

\section{Conclusions and Outlook}

In the present talk we have given the semiclassical and the algebraic
description for symmetry preserving D-branes along the line of exact
marginal deformations of the $SU(2)$ WZW model. The presentation has
been rather short, more details on the semiclassical description can
be found in\cite{Forste:2001gn}, whereas the algebraic approach will
be discussed in\cite{Hope}. Ref.\cite{Hope} will contain more results
than presented here including symmetry breaking branes and
generalizations beyond the $SU(2)$ example. More examples for
interesting questions in the present context concern: Flows under relevant
perturbations (see e.g.\cite{Recknagel:2000ri,Fredenhagen:2001nc}),
the inclusion of marginal boundary perturbations, extension to
non-compact groups (see e.g.\cite{Forste:1994wp,Manvelyan:2002xx}) and
many others.

{\bf Acknowledgement}  

\noindent 
I thank the organizers of the conference for the kind invitation to
present this talk and for creating a very pleasant and stimulating
atmosphere.\\ 
This work is supported in part by the European
Community's Human Potential 
Programme under contracts HPRN--CT--2000--00131 Quantum Spacetime,
HPRN--CT--2000--00148 Physics Across the Present Energy Frontier
and HPRN--CT--2000--00152 Supersymmetry and the Early Universe, and
INTAS 00-561.

\end{document}